\documentclass[aps,prl,reprint,superscriptaddress]{revtex4-1}

\usepackage{graphicx}
\usepackage{bm}

\renewcommand{\vec}[1]{{\mathbf #1}}

\begin{document}

\title{Absence of Anderson localization of light in a random ensemble of point scatterers}

\author{S.E. Skipetrov}
\email[]{Sergey.Skipetrov@grenoble.cnrs.fr}
\affiliation{Universit\'{e} Grenoble 1/CNRS, LPMMC UMR 5493, B.P. 166, 38042 Grenoble, France}

\author{I.M. Sokolov}
\email[]{ims@is12093.spb.edu}
\affiliation{Department of Theoretical Physics, State Polytechnic University, 195251 St. Petersburg, Russia}

\date{\today}

\begin{abstract}
As discovered by Philip Anderson in 1958, strong disorder can block propagation of waves and lead to the localization of wave-like excitations in space. Anderson localization of light is particularly exciting in view of its possible applications for random lasing or quantum information processing. We show that, surprisingly, Anderson localization of light cannot be achieved in a random three-dimensional ensemble of point scattering centers that is the simplest and widespread model to study the multiple scattering of waves.
Localization is recovered if the vector character of light is neglected. This shows that, at least for point scatterers, the polarization of light plays an important role in the Anderson localization problem.
\end{abstract}

\maketitle

Anderson localization---the appearance and dominance of localized states in strongly disordered systems---is believed to be a universal phenomenon for all quantum and classical waves \cite{anderson58,lagendijk09,abrahams10}. In particular, three-dimensional (3D) disordered systems are expected to exhibit a transition from the ``metallic'' phase with extended states to the ``insulating'' one with localized states, upon increasing the disorder \cite{evers08}. This transition was observed for electrons in disordered solids \cite{dynes10}, ultrasound \cite{hu08}, and cold atoms \cite{chabe08,kondov11,jendr12}. Reports of Anderson localization of light in 3D also exist \cite{wiersma97,storzer06,sperling13}. Here we present a theoretical study of light scattering in a 3D ensemble of resonant point scatterers (atoms) at random positions. We show that Anderson localization takes place only in the scalar approximation and disappears when the vector character of light is taken into account.
Our results raise the issue of the role that polarization effects play in the problem of Anderson localization of light in general. They suggest that it might be important to better understand these effects in more complex photonic media used in experiments: semiconductor \cite{wiersma97,beek12} or dielectric \cite{storzer06,sperling13} powders, porous semiconductors \cite{shuur99}, or disordered photonic crystals \cite{douglass11}.

The point-scatterer model is useful to understand the generic behavior of waves in disordered media \cite{lagendijk96,devries98}. In addition, this model is excellent for ensembles of cold atoms which therefore provide a fantastic and practically realizable playground for testing the theory \cite{kaiser99}.
Let us apply the point-scatterer model to study Anderson localization of light and try to go as far as possible without additional approximations. For concreteness, we assume that the point scatterers are immobile two-level atoms each having a non-degenerate ground state
$|g_i\rangle$ with energy $E_g$ and the total angular momentum $J_g = 0$ and an excited state
$|e_i\rangle$ with $E_e = E_g + \hbar \omega_0$, $J_e = 1$, and lifetime $1/\Gamma_0$ ($\hbar$ is the Planck's constant
and the index $i = 1,\ldots,N$ denotes quantities corresponding to the atom $i$ among $N$ atoms). The excited state is thus triply degenerate and splits in 3 sub-states
$\left|e_{im} \right>$ with different projections $m = -1$, 0, 1 of the angular momentum $\vec{J}_e$ on the quantization axis $z$.
The system is described by a standard Hamiltonian \cite{cohen92,morice95}
\begin{eqnarray}
{\hat H} &=& \sum\limits_{i=1}^{N} \sum\limits_{m=-1}^{1} \hbar \omega_0 | e_{im} \rangle
\langle e_{im}| +
\sum\limits_{\mathbf{s} \perp \mathbf{k}} \hbar ck
\left( {\hat a}_{\mathbf{k} \mathbf{s}}^{\dagger} {\hat a}_{\mathbf{k}\mathbf{s}} + \frac12 \right)
\nonumber \\
&-& \sum\limits_{i=1}^{N} {\hat \mathbf{D}}_i \cdot {\hat \mathbf{E}}(\mathbf{r}_i) + \frac{1}{2 \epsilon_0}
\sum\limits_{i \ne j}^{N} {\hat \mathbf{D}}_i \cdot {\hat \mathbf{D}}_j \delta(\mathbf{r}_i - \mathbf{r}_j),
\label{ham}
\end{eqnarray}
where the first two terms correspond to noninteracting atoms and the free electromagnetic field, respectively, the third term describes the interaction between the atoms and the field in the dipole approximation, and the last, contact term ensures correct description of the electromagnetic field radiated by the atoms \cite{cohen92}. Here ${\hat a}_{\mathbf{k}\mathbf{s}}^{\dagger}$ and ${\hat a}_{\mathbf{k}\mathbf{s}}$ are operators of creation and annihilation of a photon having the wave vector $\mathbf{k}$ and the polarization $\mathbf{s}$, $c$ is the speed of light in free space, ${\hat \mathbf{D}}_i$ is the dipole operator of the atom $i$, and ${\hat \mathbf{E}}(\mathbf{r}_i)$ is the electric displacement vector divided by the vacuum permittivity $\epsilon_0$ at the position $\mathbf{r}_i$ of the atom $i$. Formally solving Heisenberg equations of motion for ${\hat a}_{\mathbf{k}s}$, substituting the solution into equations for atomic operators, and applying the so-called polar approximation (i.e. neglecting retardation effects \cite{miloni74}), one obtains a system of equations for the latter operators only, with the coupling between atoms described by the so-called ``Green's matrix'' $G$ \cite{fofanov11,sokolov11,goetschy11b}. It is essentially built up of Green's functions of Maxwell equations, describing propagation of light from one atom to another. $G$ is a $3N \times 3N$ random matrix of which a particular realization is determined by the ensemble of random positions $\{ \vec{r}_i \}$ of $N$ atoms in 3D Euclidean space \cite{fofanov11,sokolov11}:
\begin{eqnarray}
G_{e_{i m} e_{j m'}} &=& \mathrm{i} \delta_{e_{i m} e_{j m'}} -
\frac{2}{\Gamma_0} (1 - \delta_{e_{i m} e_{j m'}})
\nonumber \\
&\times&
\sum\limits_{\mu, \nu}
{d}_{e_{i m} g_i}^{\mu} {d}_{g_j e_{j m'}}^{\nu}
\frac{e^{\mathrm{i} k_0 r_{ij}}}{\hbar r_{ij}^3}
\nonumber
\\
&\times& \left\{
\vphantom{\frac{r_{ij}^{\mu} r_{ij}^{\nu}}{r_{ij}^2}}
 \delta_{\mu \nu}
\left[ 1 - \mathrm{i} k_0 r_{ij} - (k_0 r_{ij})^2 \right]
\right.
\nonumber \\
&-&\left. \frac{r_{ij}^{\mu} r_{ij}^{\nu}}{r_{ij}^2}
\left[3 - 3 \mathrm{i} k_0 r_{ij} - (k_0 r_{ij})^2 \right]
\right\}.
\label{eq:green}
\end{eqnarray}
Here $\vec{d}_{e_{i m} g_i} = \langle J_e m|{\hat \mathbf{D}}_i | J_g 0 \rangle$ is the matrix element of the dipole moment operator ${\hat \mathbf{D}}_i$,
$\vec{r}_{ij} = \vec{r}_i - \vec{r}_j$, and $k_0 = \omega_0/c$. The superscripts $\mu$ and $\nu$ denote projections of vectors on axes of the reference frame. Note that Eq.\ (\ref{eq:green}) exhibits a $1/r_{ij}^3$ singularity for $r_{ij} \to 0$ which can be related to the transverse nature of electromagnetic waves.

Any excitation of the ensemble of $N$ atoms coupled through the electromagnetic field can be expanded over eigenvectors $\bm{\psi}_n$ of the matrix $G$. The real and imaginary parts of its eigenvalues $\Lambda_n$ yield the frequencies $\omega_n = \omega_0 - (\Gamma_0/2) \mathrm{Re} \Lambda_n$ and decay rates $\Gamma_n/2 = (\Gamma_0/2) \mathrm{Im} \Lambda_n$ of the corresponding  eigenstates. $G$ is therefore the fundamental object to study in order to understand the behavior of collective excitations in the atomic ensemble. We will be interested in the spatial localization of $\bm{\psi}_n$ and will compare the properties of the matrix (\ref{eq:green}) that takes into account the vector character of light with those of its scalar approximation
\begin{eqnarray}
G_{e_i e_j} &=&
\mathrm{i} \delta_{e_i e_j} + (1- \delta_{e_i e_j}) \frac{e^{\mathrm{i} k_0 r_{ij}}}{k_0 r_{ij}}
\label{eq:greenscalar}
\end{eqnarray}
that is often used to further simplify the problem \cite{goetschy11a,goetschy11c}.
Note that Eq.\ (\ref{eq:greenscalar}) has a weaker singularity for $r_{ij} \to 0$ than Eq.\ (\ref{eq:green}).
Matrices similar to (\ref{eq:green}) and (\ref{eq:greenscalar}) were previously studied in Refs.\ \onlinecite{rusek96,pinheiro04,akkermans08}.

The vector character of light is considered to be irrelevant for the Anderson localization problem in a medium with a fluctuating dielectric constant \cite{john84,anderson85,john91}. However, our model (\ref{ham}) is fundamentally different because it does not reduce to the macroscopic Maxwell equations in the interesting regime of intermediate atomic density $\rho \sim k_0^3$. Therefore, significant differences between vector and scalar cases cannot be excluded beforehand.

\begin{figure}
\includegraphics[width=\columnwidth]{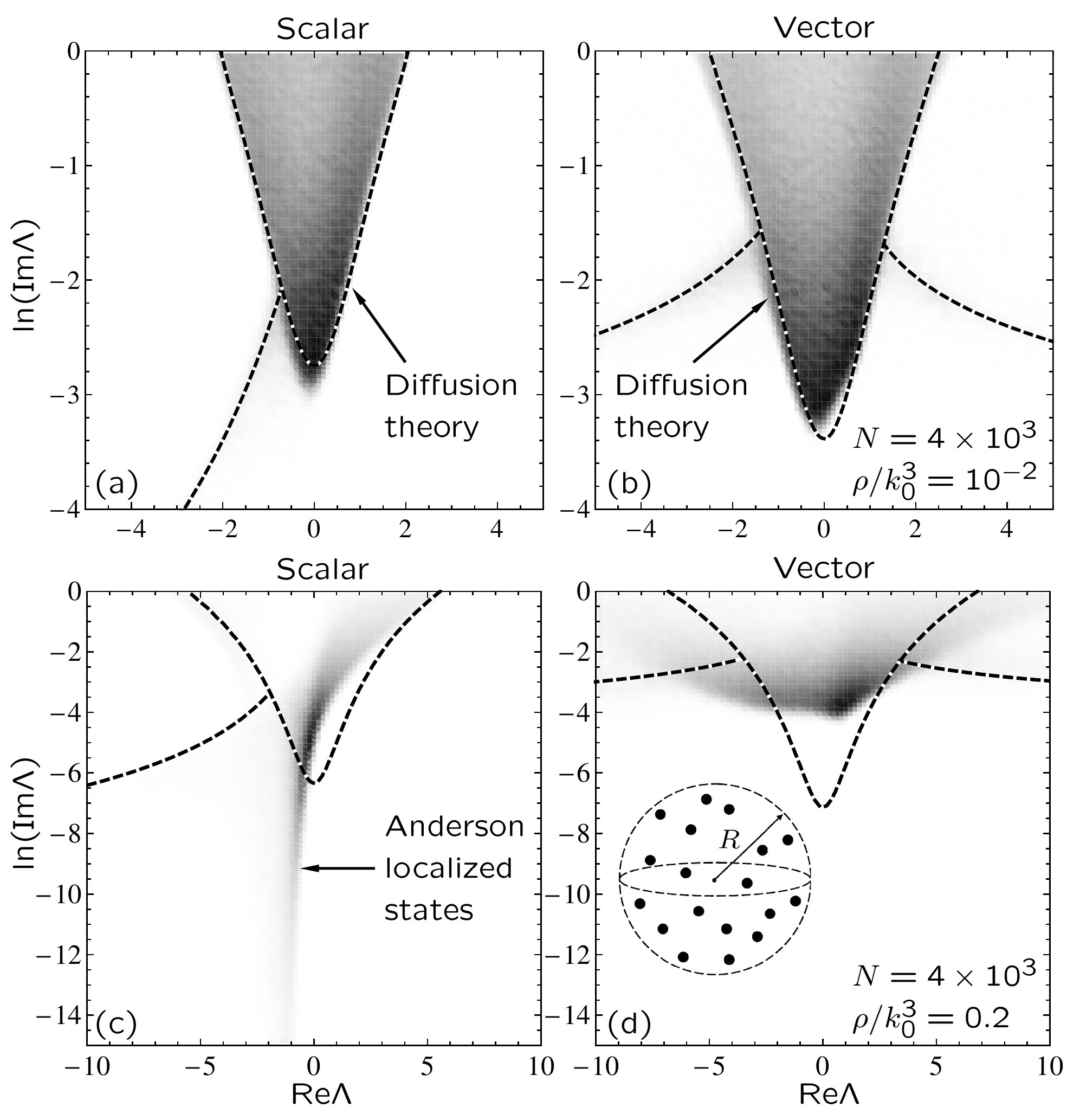}
\caption{Density of eigenvalues of the random Green's matrix. Grayscale density plots of the probability density $p(\Lambda)$ for $\Lambda$'s corresponding to long-lived states ($\mathrm{Im} \Lambda < 1$). Dashed lines show the border of the eigenvalue domain following from the diffusion theory and the spiral branches along which eigenvalues corresponding to subradiant states are concentrated \cite{goetschy11a,goetschy11b}. Panels (a) and (b) correspond to a low density of atoms at which the majority of eigenvalues are contained within the boundary imposed by the diffusion theory. Panels (c) and (d) correspond to a high density, for which states with very small decay rates $\mathrm{Im} \Lambda$ appear in the scalar model, but not in the vector one. The smallest $\mathrm{Im} \Lambda$ of the vector model is even larger than the prediction of the diffusion theory. The inset of panel (d) shows $N$ atoms (black dots) randomly distributed in a sphere of radius $R$.}
\label{fig:density}
\end{figure}

We first analyze the density of eigenvalues $\Lambda$ of the Green's matrices (\ref{eq:green}) and (\ref{eq:greenscalar}) with a particular attention paid to the part of the spectrum corresponding to long-lived states with $\mathrm{Im} \Lambda < 1$. Random realizations of Green's matrices (\ref{eq:green}) and (\ref{eq:greenscalar}) were generated by randomly choosing $N = (2 \div 8) \times 10^3$ points in a sphere of radius $R$ and volume $V$ [see the inset of Fig.\ \ref{fig:density}(d)]. Their eigenvalues $\Lambda_n$ and eigenvectors $\bm{\psi}_n = \{ \psi_{n e_{im}} \}$ obeying $G \bm{\psi}_n = \Lambda_n \bm{\psi}_n$ were computed for a sufficient number of random realizations. The density of eigenvalues $\Lambda$ for $N = 4 \times 10^3$ is shown in  Fig.\ \ref{fig:density}. At low densities $\rho = N/V$, the results obtained for scalar and vector models are similar, most of the eigenvalues being restricted to a region delimited by a line following from the diffusion theory of light scattering \cite{goetschy11a,goetschy11b}. At densities exceeding $\rho /k_0^3 \approx 0.1$, however, we observe that in the scalar model, a significant fraction of eigenvalues cross this line and acquire very small decay rates $\mathrm{Im} \Lambda$. No such long-lived states appear in the vector model.

\begin{figure}
\includegraphics[width=\columnwidth]{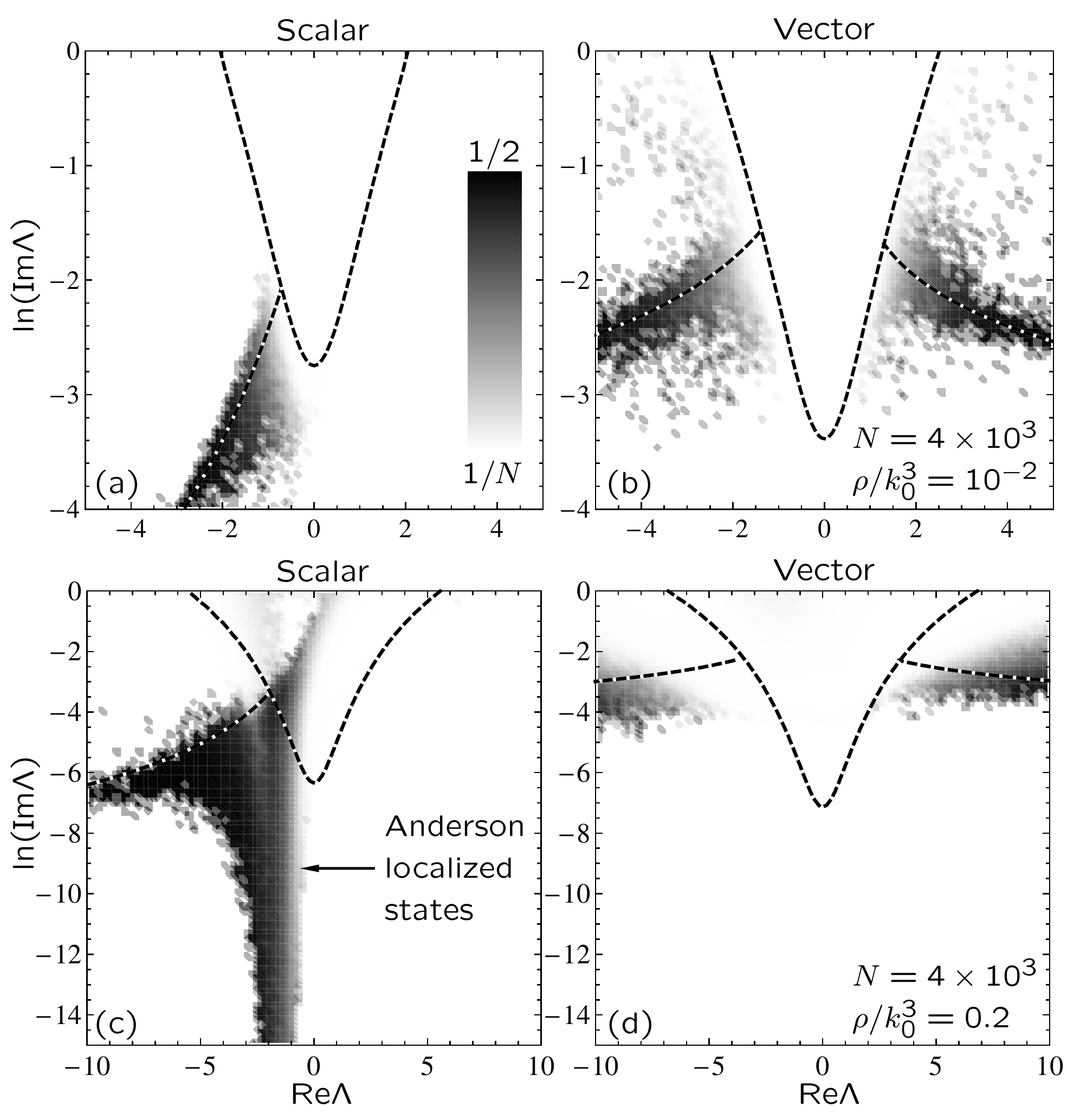}
\caption{Inverse participation ratio of eigenvectors. Grayscale density plot of the average inverse participation ratio (IPR) as a function of the eigenvalue $\Lambda$ of the corresponding eigenvector. Dashed lines are the same as in Fig.\ \ref{fig:density}. At low density, subradiant states localized on pairs of closely located scatterers exist in both scalar (a) and vector (b) models. These states have $\textrm{IPR} \simeq 1/2$. At high density, Anderson-localized states with large IPR appear in the scalar model (c), but not in the vector one (d).}
\label{fig:ipr}
\end{figure}

To test the intuitive conjecture that the long-lived states corresponding to eigenvalues with small imaginary parts may be localized in space, we show in Fig.\ \ref{fig:ipr} maps of the average inverse participation ratio (IPR) for the same parameters as in Fig.\ \ref{fig:density}. $\mathrm{IPR}_n = \sum_{i=1}^N |\psi_{n e_i}|^4/$ $(\sum_{i=1}^N |\psi_{n e_i}|^2)^2$ quantifies the degree of spatial localization of the eigenvector $\bm{\psi}_n$. It is of order $1/M$ for an eigenvector localized on $M$ atoms. In the vector model, each $\psi_{n e_i}$ is a vector with 3 components $\psi_{n e_{im}}$ and $|\psi_{n e_i}|$ should be understood as its length. As we see from Fig.\ \ref{fig:ipr}, states localized on a small number of atoms exist even at small densities. They are typically localized on pairs of very closely located atoms and are due to the phenomenon of subradiance that does not require multiple scattering and therefore has nothing to do with Anderson localization \cite{goetschy11b,goetschy11a,goetschy11c}. Their eigenvalues are concentrated along the dashed lines that depict the evolution of the smallest eigenvalue of a $2 \times 2$ Green's matrix as the distance between the two atoms is varied. In the scalar model, however, localized states of a different type appear at densities larger than $\rho /k_0^3 \approx 0.1$. These states have very small decay rates, in agreement with Fig.\ \ref{fig:density}. Once again, no such localized states are seen in the vector case.

\begin{figure}
\includegraphics[width=0.95\columnwidth]{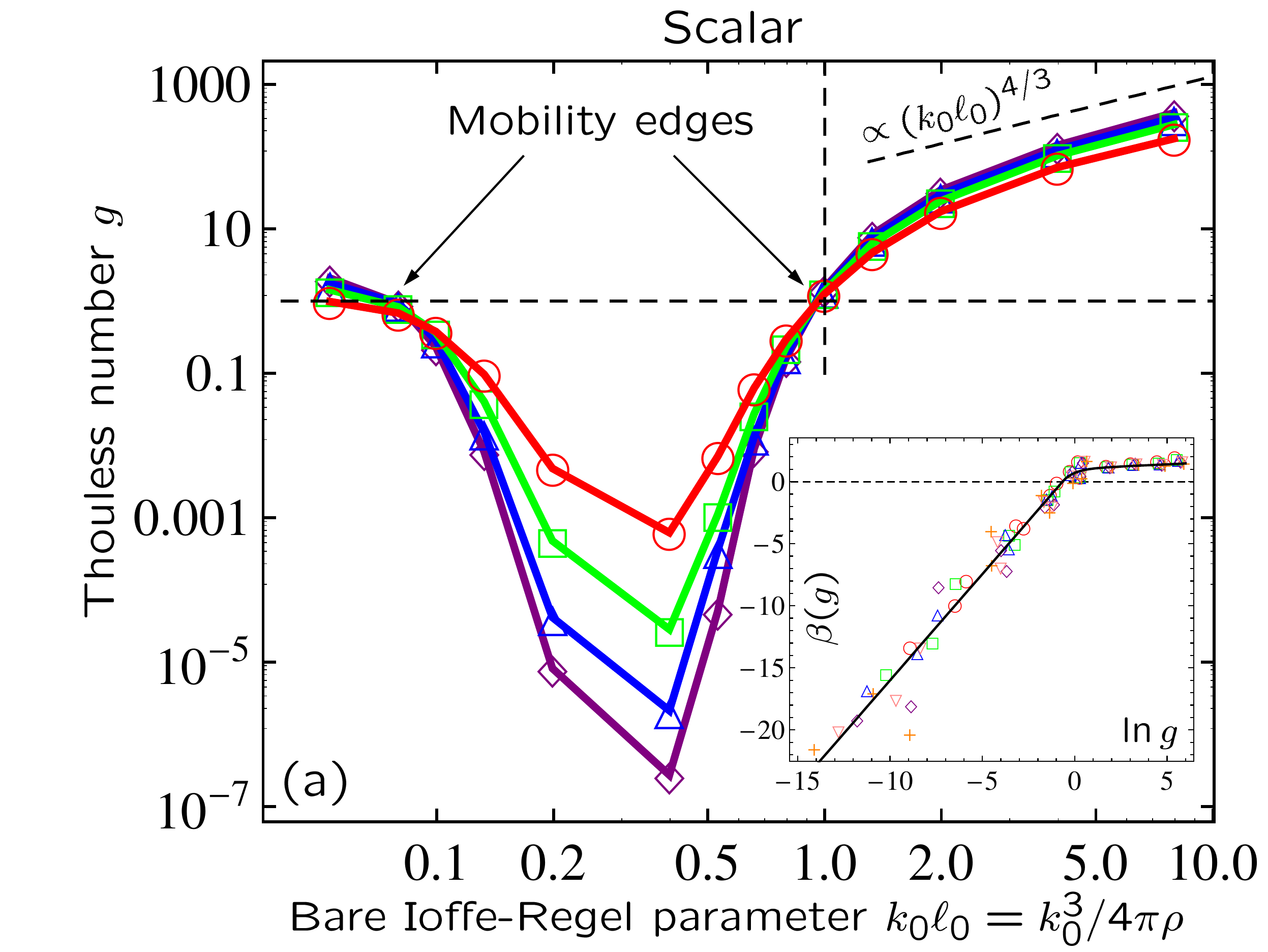}\\
\includegraphics[width=0.95\columnwidth]{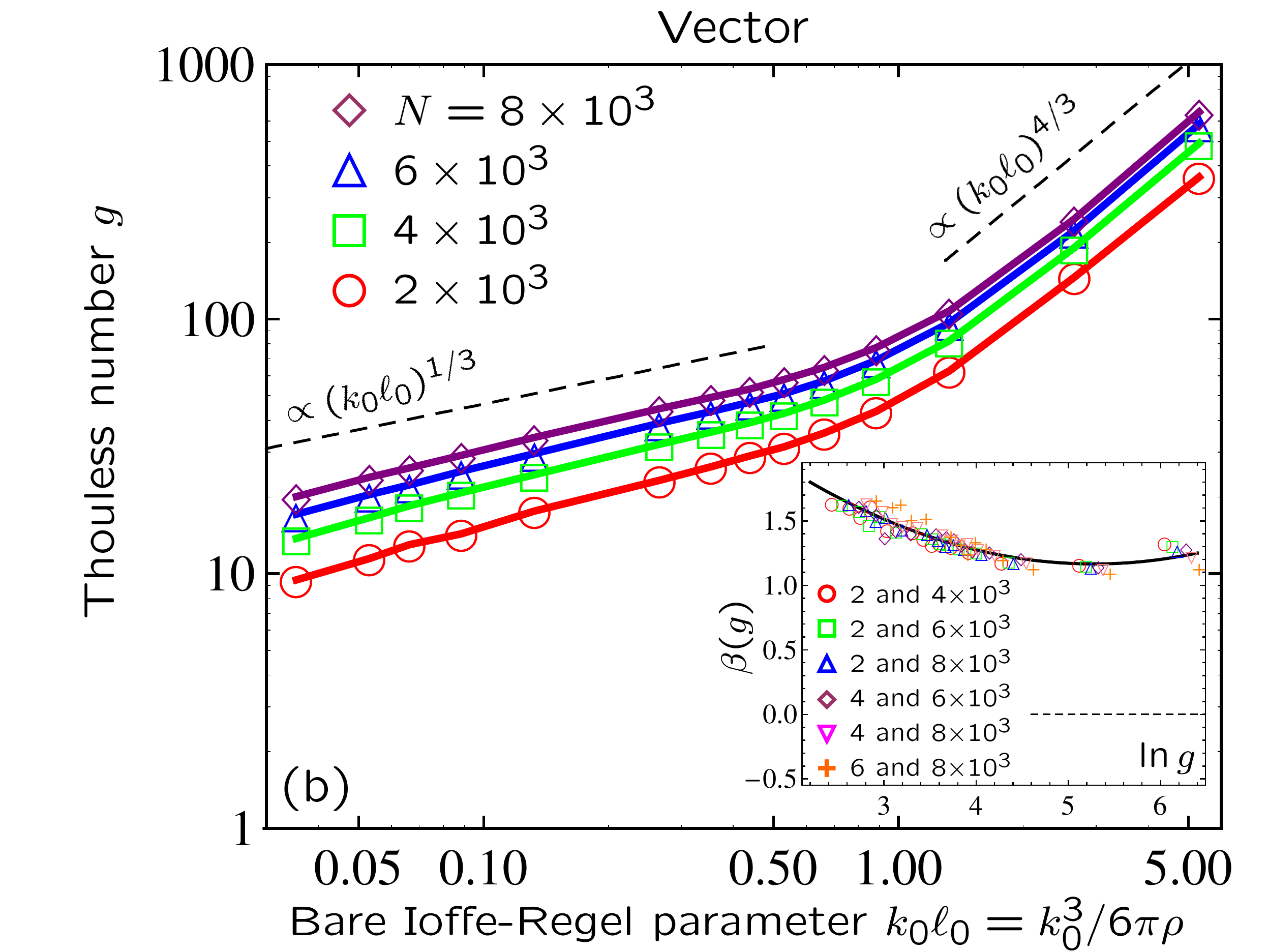}
\caption{Scaling in scalar and vector models. (a) Thouless number $g$ as a function of the bare Ioffe-Regel parameter $k_0 \ell_0$ for the scalar model at frequency $\omega = \omega_0 + \Gamma_0/2$ and for different $N$. The curves cross at $g \approx k_0 \ell_0 \approx 1$ and then again at $k_0 \ell_0 \ll 1$ and $g \approx 1$. Localization transitions take place at these points, as confirmed by the analysis of the scaling function $\beta(g)$ that changes sign at $g \approx 1$ (inset). (b) The same for the vector model. Solid lines in the insets are guides for the eye. }
\label{fig:scaling}
\end{figure}

To convince ourselves that the localized states appearing at large densities in the scalar model are due to Anderson localization, we perform the scaling analysis \cite{abrahams79}. We compute the average dimensionless lifetime of eigenstates $\langle 1/\mathrm{Im} \Lambda \rangle = \delta \omega^{-1}$ and the average spacing of nearest dimensionless eigenfrequencies $\Delta \omega = \langle \mathrm{Re} \Lambda_{n} - \mathrm{Re} \Lambda_{n-1} \rangle$ for eigenvalues $\Lambda_n$ in a strip of unit width around $\mathrm{Re} \Lambda = -1$ where, according to Figs.\ \ref{fig:density} and \ref{fig:ipr}, the localization effects are important in the scalar model. The localization transition for light at frequency $\omega = \omega_0 + \Gamma_0/2$ is expected to take place when the Thouless number (also called dimensionless conductance) $g = \delta \omega/\Delta \omega$ becomes of order unity \cite{abrahams79,wang11}. In Fig.\ \ref{fig:scaling} we show $g$ as a function of the bare Ioffe-Regel parameter $k_0 \ell_0$, with the on-resonance mean free path $\ell_0$ calculated in the independent-scattering approximation (ISA) \cite{lagendijk96}. In the scalar case, the curves $g(k_0 \ell_0)$ corresponding to different $N$ cross at $g \approx 1$, $k_0 \ell_0 \approx 1$, as expected from the Thouless and Ioffe-Regel criteria of localization \cite{evers08}. A second crossing takes place at much smaller $k_0 \ell_0$ (corresponding to a very large density $\rho$ at which ISA is not a good approximation for the mean free path $\ell$) and signals the disappearance of localization; the system starts to approach the effective medium regime. The closeness of the latter is manifest in the tendency of eigenvalues $\Lambda$ with large imaginary parts to concentrate around points on the complex plane that correspond to quasi-modes of a homogeneous sphere with some effective refractive index \cite{goetschy11a}; such a tendency is observed in both the scalar and vector models. A finite-difference estimate of the scaling function $\beta(g) = \partial \ln g/\partial \ln k_0 R$ obtained from all possible pairs of curves of the main plot is shown in the inset of Fig.\ \ref{fig:scaling}(a). As expected, $\beta(g)$ changes sign at $g \approx 1$, confirming Anderson transition in the scalar model. However, none of the above signatures of Anderson localization is seen in Fig.\ \ref{fig:scaling}(b) where we present the results for the vector model. $g(k_0 \ell_0)$ corresponding to different $N$ do not cross, $g$ always remains larger than 1, and $\beta(g) > 0$ does not change sign, suggesting no localization transition.

Let us now elucidate the reasons that prevent Anderson localization in the vector model. In the scalar approximation, the behavior of our system is analogous to that of a system of spinless fermions and $\beta(g)$ exhibits the behavior expected for the orthogonal symmetry class (see Ref.\ \cite{evers08} for a summary of the symmetry classification of disordered Hamiltonians). However, because the polarization of electromagnetic waves does not play exactly the same role as the spin of electrons, no direct analogy can be drawn between vector electromagnetic waves and the well-studied case of disordered fermionic systems.
In the system described by Eq.\ (\ref{ham}), the propagation of elementary excitations from one atom to another can be mediated not only by the transverse electromagnetic waves but also by the direct interaction of atomic dipole moments which is accounted for by the longitudinal component of the electromagnetic field \cite{lagendijk96}. The latter phenomenon becomes more and more efficient as the typical distance between neighboring atoms decreases with increasing the number density of atoms.
The possible importance of resonant dipole-dipole interactions in the context of Anderson localization was pointed out by Sajeev John in the reply \cite{john92} to a comment on his paper \cite{john91}. Later on, Nieuwenhuizen \textit{et al.} developed a perturbational approach to show that the dipole-dipole interactions between atoms in a dilute cloud of two-level atoms yield a small positive correction to the photon diffusion coefficient $D$ and thus compete with the weak localization phenomenon that tends to decrease $D$ \cite{nieuwen94}. However, being limited to low densities of atoms $\rho/k_0^3 \ll 1$, this result does not allow one to draw any conclusions concerning the fate of energy transport in the interesting regime of high atomic densities. Our calculations go beyond the perturbation theory of Ref.\ \cite{nieuwen94} and show that at high densities $\rho/k_0^3 \gtrsim 1$, the resonant dipole-dipole interactions become sufficiently strong to overcome the suppression of transport due to Anderson localization effects and thereby prevent spatial localization of elementary excitations in the system described by Eq.\ (\ref{ham}).
The dipole-dipole interactions are discarded in the scalar approximation (\ref{eq:greenscalar}) which explains essential differences between vector and scalar models. It is interesting to note that the vector character of a wave does not suppress Anderson localization if it is not accompanied by significant modifications of the near-field behavior. To demonstrate this, we repeated the calculations presented above for elastic waves which, in contrast to the electromagnetic case, can also have a propagating longitudinal component. The elastic Green's function exhibits the same $1/r_{ij}$ divergence for $r_{ij} \to 0$ as the scalar one (\ref{eq:greenscalar}) and our calculations show clear signatures of Anderson localization transition, similar to the scalar case \cite{skip13}.

At low densities $\rho/k_0^3 \ll 1$, the photon-mediated transport dominates and the dimensionless conductance of a disordered system of size $R$ is $g \propto M \ell/R$, where $M \propto (k_0 R)^2$ is the number of transport channels \cite{lagendijk96}. Assuming $\ell = \ell_0$ and noticing that $R \propto (N/\rho)^{1/3}$, we obtain $g \propto (k_0 \ell_0)^{4/3}$ at a constant $N$. As can be seen in Fig.\ \ref{fig:scaling}, this scaling is indeed obeyed at $k_0 \ell_0 > 1$ for both scalar and vector models, confirming the transport of energy via the multiple scattering of photons. At higher densities, corresponding to $k_0 \ell_0 < 1$, Anderson localization suppresses transport in the scalar model, leading to very small values of $g$, whereas the non-radiative transport channel takes over in the vector model. As follows from the approximate scaling $g \propto (k_0 \ell_0)^{1/3}$ observed in Fig.\ \ref{fig:scaling}(b), the mean free path $\ell$ is essentially independent of density $\rho$ in this regime.

Our discovery of the absence of Anderson localization of light in a 3D random ensemble of point scatterers shows that
clouds of randomly distributed cold atoms---for which the Hamiltonian (\ref{ham}) applies provided that the dipole approximation for light-matter interaction is acceptable,--- are not suitable for observation of this phenomenon.
We demonstrated the importance of the vector character of electromagnetic waves in the context of the Anderson localization problem and elucidated the role of resonant dipole-dipole interactions in multiple light scattering. In addition, our results suggest that the simple point-scatterer model might not be suitable for description of multiple light scattering in complex photonic media like, for example, the media used in recent experiments \cite{wiersma97,beek12,storzer06,sperling13,shuur99,douglass11}.
However, the latter can be modeled by grouping many point scattering centers in clusters representing large dielectric particles which, in their turn, can be distributed in space randomly or with certain spatial correlations. The role of order \cite{john87} and long- or short-range correlations \cite{rojas04} in scatterer positions can also be studied in the framework of the approach developed in this work.

\acknowledgements
SES thanks A. Goetschy and B.A. van Tiggelen for fruitful discussions. This work was supported by the Federal Program for Scientific and Scientific-Pedagogical Personnel of Innovative Russia for 2009--2013 (contract No. 14.B37.21.1938).

\end{document}